\documentclass[twocolumn,showpacs,preprintnumbers,amssymb]{revtex4}
\usepackage{graphicx}
\usepackage{dcolumn}
\usepackage{bm}
\begin{document}

\title{Eavesdropping Attack with Hong-Ou-Mandel Interferometer and
       Random Basis Shuffling in Quantum Key Distribution}

\author{Chil-Min Kim}
\email{chmkim@mail.pcu.ac.kr}
\affiliation{National Creative Research Initiative Center for Controlling
 Optical Chaos, Pai-Chai University, Daejeon 302-735, Korea}

\author{Yun Jin Choi}
\author{Young-Jai Park}
\email{yjpark@sogang.ac.kr}
\affiliation{Department of Physics and Center for Quantum Spacetime,
 Sogang University, Seoul 121-742, Korea}

\begin{abstract}
  We introduce new sophisticated attacks with a Hong-Ou-Mandel interferometer
 against quantum key distribution (QKD) and propose a new QKD protocol
 grafted with random basis shuffling to block up those attacks. When the
 polarization basis is randomly and independently shuffled by sender and
 receiver, the new protocol can overcome the attacks even for not-so-weak
 coherent pulses. We estimate the number of photons to guarantee the security
 of the protocol.

\end{abstract}

\maketitle
  A cryptography based on quantum mechanics has received much attention since
 the seminal works on quantum key distribution (QKD) by Bennett and Brassard
 (BB84) \cite{BB84} and Ekert \cite{E91}. Up to now, various QKD protocols
 have been proposed \cite{Revw,B92,Curty,Gros,Bost,Deng} and experimentally
 realized \cite{Gros,Kurt,Waks}. Also their security was continuously
 examined \cite{BB84,E91,Curty,Gros,Woj,Maye}. Recently, single photon
 QKD \cite{BB84} and entangled-state QKD \cite{E91} were much studied
 because when one does not use a single photon most protocols have their own
 serious security holes against such eavesdropping attacks as photon number
 splitting (PNS) \cite{PNS}, intercept and resend (IAR) \cite{IAR}, and
 impersonation attack \cite{Imper}. However, single photon QKD is not
 economical because it is difficult to have a reliable single-photon source
 and also a photon can be easily lost due to imperfect channel efficiency
 \cite{Agra}. For this reason, the development of a secure QKD protocol with
 not-so weak coherent pulses is indispensable to real communication.

  Very recently, two new QKD protocols that use not-so-weak coherent pulses
 (faint laser pulse) were proposed; One is based on a two-way communication
 without entanglement (LM protocol) \cite{Lucam} and the other a three-way
 communication with blind polarization \cite{KKKP}. In the former, in brief,
 the user ``Bob" prepares a qubit in one of the four states of Pauli operators
 $X$ and $Z$, and sends it to his counterpart ``Alice." With probability $c$,
 Alice measures the prepared state and, with probability $1-c$, she uses it
 to encode the message. She sends the qubit back to Bob. Then Bob can
 deterministically decode Alice's message by measuring the qubit in the same
 basis he prepared it.

  In the latter, Alice sends two randomly and independently polarized
 not-so-weak coherent pulses to Bob. Bob rotates the polarization of pulses
 with another random angle, shuffles it with $\pm \frac{\pi}{4}$ or
 $\mp \frac{\pi}{4}$, and sends back the pulses to Alice. Alice compensates
 her random angles, encodes a key bit, and sends one of the pulses to Bob
 after randomly blocking the other. Then Bob reads the polarization of the
 return qubit after compensating his random angle. When Alice publicly
 announces the blocking factor, Bob recovers the key bit.

  The security of the former protocol was examined in a noisy channel against
 a spy pulse. And it was claimed that the protocol is robust against the PNS
 attack because of a lack of symmetry in the photon states. In the latter,
 the security of the random polarization was examined against the PNS and the
 IAR attacks. And though it was expected that the shuffling and random
 blocking would play a crucial role in enhancing security, the protocol turned
 out insecure, particularly against the impersonation attack \cite{Zhang}.
 So Kye and Kim modified the protocol by randomly and independently shuffling
 the qubit polarization with $\frac{\pi}{4}$ or $-\frac{\pi}{4}$ (KK protocol)
 \cite{Kye-Kim}.

  However, we are still doubtful about the security of both the LM and the KK
 protocols. To use them in practice, the security must be rigorously examined.
 So we develop new sophisticated eavesdropping attacks using a Hong-Ou-Mandel
 interferometer (HOMI) \cite{HOM}, which are the most advanced ones against
 these type QKD protocols. In this Letter, first, we introduce the new attacks
 to show the security holes of the LM and the KK protocols. Next, we propose
 a new QKD protocol that uses not-so-weak coherent pulses. Last, we prove
 the security of our protocol against the attacks that we introduce.

  We introduce the PNS attack with a HOMI to examine the security of the LM
 protocol. The attack procedure is like this. When a not-so-weak coherent
 pulse is used in a lossy channel, an eavesdropper Eve replaces the lossy
 channel with a perfect one and splits out photons from the forward and the
 backward path. Then Eve measures the interference between the split photons
 from both the paths with a HOMI. If interference appears, the coding is "0";
 if not, it is "1." Thus Eve obtains the key bit regardless of the lack of
 symmetry.

  For the security of the KK protocol, we now introduce a new impersonation
 attack with a HOMI. When Eve has a HOMI in her superiority, she can easily
 attack the protocol even though the shuffling method is modified to block up
 an impersonation attack. The procedure is as follows: (1) Eve intercepts the
 two qubits $|\psi_1\rangle=|\theta_1\rangle \otimes |\theta_2 \rangle$ from
 Alice to Bob, and stores them. Then Eve prepares two highly coherent qubits
 $|\psi'_1 \rangle = |\theta'_1\rangle \otimes |\theta'_2 \rangle$, and sends
 them to Bob. (2) When the qubits are back from Bob, Eve compensates her random
 angles (let the compensated qubits be $|\Psi\rangle$), splits out one photon
 from both qubits of $|\Psi \rangle$ and measures the angle difference with a
 HOMI. Because of the random and independent shuffling $\pm \frac{\pi}{4}$, the
 qubits in $|\Psi \rangle$ are either parallel or orthogonal: if interference
 occurs, the two qubit states are parallel; if not, they are orthogonal. When
 they are parallel, Eve applies
 $\hat{U}_y (\frac{\pi}{4}) \otimes \hat{U}_y (\frac{\pi}{4})$ to
 $|\psi_1 \rangle$; if not, she applies $\hat{U}_y(-\frac{\pi}{4}) \otimes
 \hat{U}_y (\frac{\pi}{4})$. She sends the qubits to Alice. (3) Eve measures
 the pre-key bit after intercepting the return qubit from Alice, and estimates
 the key bit according to the blocking factor. She applies the estimated key
 bit to one of the qubits of $|\Psi \rangle$ depending on the blocking factor
 and sends the chosen qubit to Bob. (4) When Alice publicly announces the
 blocking factor, Eve recovers the key bit.

  In this attack, let us consider the case that the two qubits in $|\Psi
 \rangle$ are parallel. In (3), the qubit state, in Eve's measurement, is
 either $|0 \rangle$ or $|\frac{\pi}{2}\rangle$, since Eve applies $\hat{U}_y
 (\frac{\pi}{4}) \otimes \hat{U}_y (\frac{\pi}{4})$ to $|\psi_1 \rangle$.
 Then Eve obtains the key bit regardless of the blocking factor. After the
 measurement, Eve applies ${\hat U}_y ((-1)^k \frac{\pi}{4})$ to any of the
 qubits in $|\Psi \rangle$ depending on her measurement, and sends it to Bob
 without revealing her presence in the channel. When the two qubits in
 $|\Psi \rangle$ are orthogonal, Zhang's attack protocol \cite{Zhang} is
 valid. Thus Eve can attack the KK protocol perfectly.

{\it Protocol}. --- To block up the impersonation attack with a HOMI and
 to use not-so-weak coherent pulses, we adopt the basic idea of the BB84
 protocol, which is to use the four photon states of $0$, $\frac{\pi}{2}$,
 and $\pm \frac{\pi}{4}$ polarization. The four states can be written as
 $(-1)^s \frac{\pi}{4}+ \{(-1)^r+1\} \frac{\pi}{8}$, where $s$ is the
 random polarization shuffling and $r$ is the random basis shuffling. Here
 the basis shuffling plays a crucial role in blocking up the impersonation
 attack. Our new protocol with random basis shuffling proceeds as follows:

\begin{itemize}

\item[(P.1)] Alice sends two qubits of $|\psi_1\rangle = |\theta_1 \rangle
  \otimes |\theta_2\rangle \equiv \bigotimes_{b=1}^2 |\theta_b\rangle$
  to Bob.

\item[(P.2)] After receiving $|\psi_1\rangle$, Bob applies a unitary operator
  $\bigotimes_{b=1}^2 \hat{U}_y(\phi +(-1)^{s_b} \frac{\pi}{4}
  + \{(-1)^{r_b}+1\} \frac{\pi}{8})$ where $s_b=\{0, 1\}$ and $r_b=\{0, 1\}$
  are the independent random numbers to shuffle the photon state and the
  polarization basis, respectively. He returns the qubits $|\psi_2 \rangle$
  to Alice.

\item[(P.3)] On receiving $|\psi_2 \rangle$, Alice applies $\bigotimes_{b=1}^2
  \hat{U}_y (-\theta_b + (-1)^{k_b} \frac{\pi}{4} + \{(-1)^{p_b}+1\}
  \frac{\pi}{8})$, where $k_b \in \{0, 1\}$ is the key bit and
  $p_b \in \{0, 1\}$ is Alice's basis shuffling parameter. She block one
  of the qubits and sends the other $|\psi_3 \rangle$ to Bob.

\item[(P.4)] When $|\psi_3 \rangle$ arrives, Bob compensates his random angle
  with $-\phi$, divides the qubit $|\psi_3 \rangle$ into two with a $50$
  percent beam splitter, and measures each pre-key bit on the
  $|\pm \frac{\pi}{4}\rangle$ and the $|0 \rangle$ and
  $\frac{\pi}{2} \rangle$ bases. He stores the pre-key bit.

\item[(P.5)] After repeating the procedure from (M.1) to (M.4) $N$-times, Alice
  publicly announces $b$ and $p_b$. Then Bob decodes the original key bit.

\item[(P.6)] When Eve misses the key bit because of the division of the return
  qubit, Bob publicly announces on which turns qubits have been missed in
  measurement. Then Alice and Bob repeat the procedure from (M.1) to (M.5)
  for the missed key bit until the full key bit stream is generated.

\item[(P.7)] In order to verify the integrity of the shared keys, Alice and
  Bob evaluate the hash values, $h_a = H(k_a)$ and $h_b = H(k_b)$, where
  $k_a$ and $k_b$ are Alice's and Bob's shared keys, respectively. Then they
  exchange and compare them. If $h_a = h_b$, they keep the shared keys,
  otherwise, they abolish the keys.
\end{itemize}

  In this protocol, the efficiency of key distribution depends on the number
 of photons of $|\psi_3 \rangle$. The efficiency is $1-1/2^n$ for an
 $n$-photon qubit. If one wants to increase the efficiency, (s)he can
 slightly modify the protocol like this. In (M.4) Bob stores the return
 qubit $|\psi_3 \rangle$ in a quantum storage like a fiber and publicly
 announces to Alice his reception of the qubit. Then when Alice announces
 $b$ and $p_b$, Bob decodes the original key bit by measuring the polarization
 of the stored qubit.

  Now, we focus on the security against the impersonation attack, since it was
 proved that a protocol using random angle polarization is secure against the
 PNS and the IAR attacks \cite{KKKP}.

{\it Attack-1}. --- We suppose that the superior Eve knows the angle
 difference of the two qubits in $|\Psi \rangle$ and the pre-key bit exactly.
 Then the attack procedure is as follows:

  (A.1) After (P.1), Eve intercepts and stores $|\psi_1 \rangle$, and sends
 $|\psi'_1 \rangle = \bigotimes_{b=1}^2 |\theta'_b\rangle$ to Bob.

  (A.2) After (P.2), Eve intercepts $|\psi'_2 \rangle$ and compensates her
 random angle with $-\theta'_b$. Then Eve has $|\Psi \rangle =
 \bigotimes_{b=1}^2 |(\phi +(-1)^{s_b} \frac{\pi}{4} + \{(-1)^{r_b}+1\}
 \frac{\pi}{8})\rangle$. Eve splits out a few photons from both pulses
 of $|\Psi \rangle$ and stores the rest. Then Eve measures the angle
 difference of the split photons with a HOMI. There are three cases of
 results: first, on complete non-interference the angle difference of
 the two pulses is $\frac{\pi}{2}$; second, on complete interference
 it is $0$; and third, on partial interference it is $\pi/4$. On each
 case, Eve applies $\hat{U}_y(\frac{\pi}{4})\otimes \hat{U}_y(-\frac{\pi}{4})$,
 $\hat{U}_y(\frac{\pi}{4})\otimes\hat{U}_y(\frac{\pi}{4})$, and
 $\hat{U}_y(\frac{\pi}{4}) \otimes \hat{U}_y(0)$ to $|\psi_1 \rangle$,
 respectively. And Eve sends $|\psi^e_2 \rangle$ to Alice, where the
 superscript $e$ implies Eve's action to Alice's qubits.

  (A.3) After (P.3), Eve measures the pre-key bit from $|\psi^e_3 \rangle$,
 estimates Alice's unitary operation depending on $b$ from the pre-key bit,
 chooses one qubit of $|\Psi \rangle$, and applies
 the unitary operator that she has estimated.

  In order to show the security of our new protocol, we consider the case
 that Bob applies a unitary operator $\hat{U}_y(\phi) \otimes \hat{U}_y
 (\phi + \frac{\pi}{4})$. In (A.2), on receiving the qubits, Eve compensates
 her random angles. Then the qubit state becomes $|\Psi\rangle = |\phi \rangle
 \otimes |\phi + \frac{\pi}{4} \rangle$. Suppose that Eve applies $\hat{U}_y
 (\frac{\pi}{4}) \otimes \hat{U}_y(0)$ to the qubits $|\psi_1 \rangle$ and
 returns the qubits $|\psi^e_2 \rangle$ to Alice, since she does not know the
 sequence of the qubits. She can measure only the angle difference with a HOMI.
 Also suppose that, on receiving the qubits, Alice compensates her random
 angles, blocks the second qubit, rotates the first by $\frac{\pi}{4}$, and
 sends the first to Bob. Then the parameters of the qubit are $b=1$, $k_1=0$,
 and $p_1=1$. Eve intercepts the return qubit $|\psi^e_3 \rangle$ and measures
 the polarization of the qubit that is $\frac{\pi}{2}$. Here Eve must estimate
 the rotation angle depending on $b$. When Eve chooses $b=2$, the rotation
 angle is $\frac{\pi}{2}$. Then the parameters that Eve estimates are $k_2=0$
 and $p_2=0$. Eve rotates the second qubit of $|\Psi \rangle$ by
 $\frac{\pi}{2}$, and sends it to Bob. Then Bob's pre-key bit is
 $|\frac{3}{4}\pi\rangle = - \frac{\pi}{4}\rangle$. When Alice announces $b$
 and $p_b$, Bob recovers the key bit as $k=1$. When Eve chooses $b=1$,
 Bob's key bit is $k=0$. Whether the angle difference of the two qubits in
 $|\Psi \rangle$ is $\frac{\pi}{2}$ or $0$, there is no error, whatever
 the sequence of the qubits in $|\Psi \rangle$ is. Bob's wrong recovery is
 caused by Eve's wrong choice of the sequence when the polarization difference
 between the two qubits in $|\Psi \rangle$ is $\frac{\pi}{4}$. Owing to the
 possibility of the $\frac{\pi}{4}$ angle difference, sequence mismatch, and
 wrong choice of $b$, Bob's error rate is $12.5$ percent. This means our new
 protocol is secure against Eve's impersonation attack, even when she knows
 the angle difference between the two qubits in $|\Psi \rangle$ and
 the pre-key bit.

\begin{figure}
 \begin{center}
  \rotatebox[origin=c]{0}{\includegraphics[width=8.5cm]{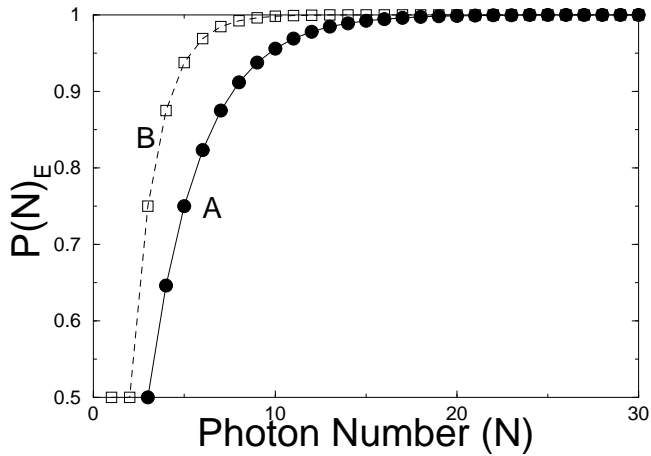}}
    \caption{The probability of Eve's estimation depending on the photon
     number: Line A is the pre-key bit estimation with the use of POVM,
     and Line B is the angle difference and sequence estimation with
     a HOMI.}
 \end{center}
\end{figure}

{\it Attack-2}. --- When Eve knows not only the pre-key bit but also the
 angle difference and the sequence of the two qubits in $|\Psi\rangle$,
 she can recover the key bit completely without Bob's recognition. However, a
 measurement of all of them is not easy in practice because of the limited
 number of photons. To estimate the number of photons for the security of our
 new protocol, first, we consider Eve's attack on the pre-key bit with the
 positive operator-valued measurement (POVM) \cite{POVM}. In this measurement,
 when we consider one of the four photon states and $N$ identical copies of
 the state, we can obtain the probability of Eve's estimation of the pre-key
 bit depending on the number of photons. According to ref.\cite{POVM}, the
 probability is $P(N)_E = 1 - (1/2)^{[(N-1)/2]}$, where $[~\cdot~]$ is the
 rounding to the closest lower integer. Line A in Fig. 1 shows about $95$
 percent accuracy for $N=10$.

  Next, we consider Eve's attack on the angle difference and the sequence
 between the two qubits in $|\Psi \rangle$. Suppose that Eve replaces the
 lossy second and third channels with perfect ones and that she has perfect
 technology to split a certain number of photons from both the qubits,
 although this is far beyond today's technology. Eve splits out the same
 number of photons from both the qubits in $|\Psi \rangle$ in consideration
 of the channel efficiency. She picks out one photon from $N$ photons split
 from the first qubit (let it be $T_1$ and the others $T_2$), and picks out
 one photon from $N$ photons split from the second qubit (let it be $R_1$
 and the others $R_2$). From the interference between $T_2$ and $R_2$, Eve
 measures the angle difference between the two qubits in $|\Psi \rangle$
 using the method in (A.2). When $T_2$ and $R_2$ give rise to partial
 interference, the angle difference is $\pi/4$. Then Eve rotates $T_1$ by
 $\pi/4$ and measures the interference between $T_1$ and $R_1$. When she
 observes interference, the angle of the first qubit to the second one is
 $-\frac{\pi}{4}$, while with no-interference it is $\frac{\pi}{4}$. Then
 Eve knows the sequence of the qubits for the $\pi/4$ angle difference. In
 the case of partial interference between $T_2$ and $R_2$, let us assume
 that $j$ photons make interference while $N-1-j$ photons give rise to no
 interference. Then the probability of Eve's estimation for the partial
 interference is $\frac{1}{2^{N-1}} \sum_{j=1}^{N-2}{{N-1} \choose j}$. For
 the other cases of complete interference and no interference, Eve regards
 that the photon states of $T_2$ and $R_2$ are parallel and orthogonal,
 respectively. Then Eve's probability for the estimation of the angle
 difference and the sequence of $|\Psi \rangle$ is
 $P(N)_E = \frac{1}{2}+ \frac{1}{2^{N}} \sum_{j=1}^{N-2} {{N-1} \choose j})$
 because of the probability of the $\pi/4$ angle difference.

  Line B in Fig. 1 is the probability of Eve's estimation for the angle
 difference and the sequence of the two qubits in $|\Psi \rangle$ depending
 on the number of split photons. When Eve splits out $5$ photons from each
 qubit she can measure both the angle difference and the sequence of the two
 qubits with about $93$ percent accuracy. Lines A and B in Fig. 1 show that
 the estimation of the pre-key bit of $|\psi^e_3 \rangle$ is less efficient
 than that of the angle difference and the sequence of the two qubits in
 $|\Psi \rangle$, even when the channel efficiency is considered. When we
 consider that the qubits of $|\Psi \rangle$ are at Eve's mercy, we can
 understand that most of the errors by Eve can occur in the measurement of
 the pre-key bit due to the basis shuffling. So the basis shuffling is
 decisive in blocking up Eve's attack.

{\it Attack-3}. --- Another instance of the impersonation attack in our new
 protocol is the attack on $b$ and the photon state of $|\psi^e_3 \rangle$.
 In this attack, Eve applies $\hat{U}_y (\frac{\pi}{8}) \otimes \hat{U}_y(0)$
 to $|\psi_1 \rangle$ in (A.2) and sends $|\psi^e_2 \rangle$ to Alice. Since
 Alice applies a key bit and basis shuffling, after she compensates her random
 angle with $-\theta_b$, the qubit state $|\psi^e_3 \rangle$ in (A.3) is one
 of the four states $(1+2n)\pi/8$ for the first qubit or one of the four
 states $2n \pi/8$ for the second, where $n=0, 1, 2, 3$. Then by measuring
 the qubit state with POVM, Eve can obtain $b$ and $k \oplus p_b$. Depending
 on $b$, Eve applies $k \oplus p_b$ to the $b$-th qubit of $|\Psi \rangle$ and
 sends the qubit to Bob. When Alice publicly announces $b$ and $p_b$, Eve
 recovers the key bit.

  In this attack protocol, Eve should measure the return qubit state with
 POVM among eight states. We can intuitively understand that the photon
 state estimation with POVM among eight states is less efficient than that
 of among four states, since POVM for eight states needs at least $7$
 photons \cite{POVM}. The attack on the pre-key bit is more serious than
 the attack on $b$. Eve can also attack $b$ by counting the number of
 photons of the two pulses \cite{Zhang}. To block up this attack, in (P.3),
 the number of the photons of the returning pulse should be randomly
 reduced to be less than either of the photon numbers of the two received
 pulses.

{\it Attack-4}. --- Eve can add an invisible spy pulse, whose wavelength is
 different from that of Alice's qubits \cite{QCai}. The removal of this spy
 pulse is so trivial when Alice uses a commercial band-pass filter, a
 spectrometer, and a Fabry-Perot interferometer. To block up this kind of
 attack, the use of quasi-monochromatic photons is crucial. For another
 instance, Eve can add a spy pulse with time delay to the original qubits.
 Alice and Bob can easily remove this spy pulse with an optical switcher.
 Alice and Bob can also recognize the spy pulse by randomly measuring the
 pulse intensity.


  In conclusion, we have shown that the LM and the KK protocols are vulnerable
 to sophisticated eavestropping attacks with a Hong-Ou-Mandel interferometer.
 The LM protocol is insecure against the PNS attack with a HOMI and the KK
 protocol against the impersonation attacks with a HOMI. These atacks are
 effective to these protocols. To overcome these attacks, we have proposed
 a new protocol with basis shuffling as an altenative. In the three-way
 communication ptotocol, when both the polarization basis and the photon
 state are randomly and independently shuffled, the protocol with
 random polarization becomes robust against not only the PNS and the IAR
 attacks but also the sophisticated impersonation attacks with a HOMI, even
 with not-so-weak coherent state pulses. As we have shown, the number of
 photons of Alice's qubits is very important in blocking up the impersonation
 attacks. This new QKD protocol can be applicable to real communication
 because of the merit of robustness and the use of not-so-weak coherent pulses.

  This work was supported by the Creative Research Initiatives of the
 Korean Ministry of Science and Technology. Y. J. Park were supported by
 the Science Research Center Program of the Korean Science and Engineering
 Foundation with grant number R11-2005-021.

\end{document}